# Terahertz Nonlinear Index Extraction via Full-Phase Analysis

Kareem J. Garriga Francis, Mervin Lim Pac Chong, Yiwen E, And X.-C. Zhang[*]

*The Institute of Optics, University of Rochester, Rochester, NY 14627, USA*

*Corresponding author: xi-cheng.zhang@rochester.edu*

**We experimentally show the spectrally averaged nonlinear refractive index and absorption coefficient for liquid water, water vapor, α-pinene, and Si using a full-phase analysis in the terahertz regime through a standard time-domain spectrometer. Our results confirm that the nonlinear index of refraction of the liquid samples in this regime exceeds the near-infrared optical nonlinear index by six orders of magnitude. In the case of liquid water and water vapor at atmospheric pressure, we find a nonlinear index of $7.8 \times 10^{-10}$ cm$^2$/W and $6 \times 10^{-11}$ cm$^2$/W respectively, which are both much larger than expected.**

As more energetic sources of terahertz (THz) radiation are made available, there is a growing interest in understanding nonlinear effects in this elusive spectral region [1-3]. Recent investigations employing Z-scan techniques have shown that some materials exhibit nonlinear indices that are nearly six orders of magnitude larger than their optical counterparts [4]. Given the results found via the Z-scan method, in the specific case of analyzing the nonlinearity of liquid water, a THz peak electric field strength exceeding 3 MV/cm is enough to modify the refractive index by one order of magnitude. This change is enough to alter focusing and change expected results from spectroscopy while the wave maintains its non-ionizing properties, leading to several interesting nonlinear phenomena such as the possibility of enacting filamentation without plasma formation or elucidating nonlinear processes without the onset of optical breakdown. As such, it becomes apparent why probing nonlinearities in the THz regime is very exciting. Unfortunately, for Z-scan to be used effectively, the beam spatial mode ($M^2$) must be well known and the source cannot be single-cycle [5]. Such constraints are not trivially solved in the optical regime and are even more cumbersome in the THz regime where the sources are broadband and spatiotemporal focusing issues are abundant. In order to mediate mode issues in the optical regime, full-phase analysis experiments have been conducted using second-harmonic frequency-resolved optical gating (SHG FROG) and pulse reconstruction yielding much more accurate values for the nonlinear index of refraction than Z-scan techniques [5].

In this letter, we show experimentally obtained values for the modification of the dielectric constants due to nonlinearity of liquid water, α-pinene, and Si. Given that THz time-domain spectroscopy (TDS) allows for the extraction of the phase of a captured signal, a full-phase analysis is instead conducted to extract nonlinear index $n_2$ and nonlinear absorption $\alpha_2$. Our results show that material nonlinearities can be characterized directly with conventional THz-TDS systems. When spectrally averaged, the index of refraction indeed matches values previously obtained in Z-scan experiments. Moreover, improvements on the technique and the detection system would allow for spectrally resolved values of the Kerr coefficients. Similar experiments have been conducted in the optical regime using second-harmonic frequency-resolved optical gating (SHG FROG) and pulse reconstruction yielding much more accurate values for the nonlinear index of refraction than Z-scan techniques [5].

All samples (liquid water, α-pinene, and Si) are evaluated via a THz-TDS system based on a two-color air-plasma source [6-7]. An amplified laser capable of providing 3 mJ, 800 nm, <50 fs pulses at a 1 kHz repetition rate is used for the experiment. The laser is split into a pump-probe configuration with 2 mJ reserved for the pump beam. The resulting THz radiation is collected and guided to a conventional free-space electro-optic (EO) sampling setup [8] and controlled with a variable attenuator placed along the path of the pump beam. The THz peak electric field is measured at the sample position with EO sampling prior to measuring the samples. The peak electric field strength can be tuned between 10 kV/cm to 230 kV/cm. Although the plasma source is essentially different and its position moves slightly when the optical pump energy is varied, these effects are accounted for in the computation by recording signals as system references. Lastly, a knife-edge test is done to verify that the THz focal spot size leads to a Rayleigh range (~ 2 cm) that is larger than the sample thickness (~ 2 mm).

The sample cells are fabricated with a 3D printer using natural white polylactic acid (PLA) material with a 100% infill and 0.06 mm layer height. The cells are treated as three-layer structures where 2-mm thick α-pinene and 350-μm thick water films are studied, respectively. In order to check for the validity of the analysis, Si is treated as a three-layer structure where two flat 500-μm thick pieces of PLA sandwich an 800-μm thick Si wafer rather than as a single layer structure. We choose samples that are known to be noncentrosymmetric and whose responses have been previously characterized with Z-scan methods in order to compare with our full-phase analysis method. In the specific case of α-pinene, this liquid is chosen because it is a non-polar liquid and it shows promise toward the development of THz liquid lasers.

In our full-phase analysis, we extract the index and absorption data for all investigated samples at various peak THz electric field strengths. We then use a fitting operation to relate these parameters to the nonlinear phase and extract the nonlinear indices. Extraction of the refractive index and absorption coefficient rely on conventional spectroscopy techniques [9-10]. The transfer function is found as the ratio of sample to reference fields in the frequency domain,

$$H(\omega) = \frac{\tau E_{sam}(\omega)}{E_{ref}(\omega)} \times FP(\omega). \quad (1)$$

In **Eqn. (1)**, τ represents the Fresnel transmission coefficient while FP(ω) represents the Fabry-Perot effects [11]. In order to determine the index of refraction of the samples within the cell in an efficient manner, the index (linear and nonlinear) of the PLA material must first be determined. We print a test flat with thickness of 2mm and use it as a sample. If a single sample is used, there is an external Fresnel transmission coefficient, an internal transmission coefficient, an external reflection, an internal reflection, and a Fabry-Perot term in **Eqn. (1)**. For the nonlinear measurements, the index of refraction is extracted at the focus for eight peak electric field values. The index and absorption coefficient are found through the relations by **Eqns. (2)** and **(3)**:

$$\phi^T(\omega) = \frac{\omega L_{sam}}{c}\left[n_{sam}(\omega) - n_{air}(\omega)\right] \quad (2)$$

and

$$\ln\left[|H(\omega)|\right]^T = -\frac{L_{sam}}{2}\left[\alpha_{sam}(\omega) - \alpha_{air}(\omega)\right]. \quad (3)$$

Here, $n_{sam}$ and $L_{sam}$ are the sample index of refraction and the sample thickness, respectively. It is expected that $n_{sam}$ and $\alpha_{sam}$ will vary slightly with increasing THz peak intensity caused by nonlinearity. The (T) superscript denotes that the operation is performed on the test flat (PLA material). The phase noise is accounted for and measured as the standard deviation across three averages for each peak electric field value probed. The index of refraction is defined as $n(\omega) = n^o(\omega) + n_2 I_{THz}$ where $n^o$ denotes the linear portion, $n_2$ is the Kerr coefficient in m$^2$/W, and $I_{THz}$ is the THz peak intensity in W/m$^2$ [12-13]. Similarly, $\alpha(\omega) = \alpha^o(\omega) + \alpha_2 I_{THz}$, where $\alpha^o$ is the linear absorption coefficient in m$^{-1}$ and $\alpha_2$ denotes the two-photon absorption in m/W. The lowest intensity scan defines the linear dielectric constants and the high electric field scans expose the trend of the index with respect to intensity.

Since the water vapor in our laboratory environment is not controlled, nor is it computationally removed, the nonlinear index of the samples cannot be decoupled from the nonlinear index of the water vapor and we must instead represent the nonlinearity as a difference between the effects of the sample material and water vapor. In our model, water vapor is the main contributor to the nonlinearity in air (the reference material) as the nitrogen, oxygen, and argon molecular modes cannot induce a non-linear response [1]. As such, treating water vapor as the reference and applying the full definition of the index of refraction, **Eqns. (2)** and **(3)** lead to:

$$\phi^T(\omega) = \frac{\omega L_{sam}}{c}\left[n_d^o(\omega) + n_{2,d}(\omega) I_{THz}\right] \quad (4)$$

and

$$\ln\left[|H(\omega)|\right]^T = -\frac{L_{sam}}{2}\left[\alpha_d^o(\omega) + \alpha_{2,d}(\omega) I_{THz}\right] \quad (5)$$

where $n_d^o(\omega) = n_{sam}^o(\omega) - n_{air}^o(\omega)$, $n_{2,d}(\omega) = n_{2,sam}(\omega) - n_{2,air}(\omega)$, and $\alpha$ parameters follow respectively. A fit can then be done to **Eqns. (4)** and **(5)** to extract $n_{2,d}$ and $\alpha_{2,d}$ for every point along the THz frequency space.

A similar process is done for determining the index of the samples within the cells. Given that the PLA index is known for different peak electric fields, the sample can be treated as a three-layer structure. Here, $n_{mat}$ now designates the known index of refraction of the PLA while $n_{sam}$ relates to the sample within the cell. There are now two internal reflection and transmission coefficients, two external reflection and transmission coefficients, and three Fabry-Perot terms to evaluate in **Eqn. (1)**. The dielectric constants are extracted as:

$$n_{sam}(\omega) = \frac{\frac{c\phi^S(\omega)}{\omega} + n_{air}(\omega) L_{eff} - n_{mat}(\omega)\left[L_{eff} - L_{sam}\right]}{L_{sam}} \quad (6)$$

and

$$\alpha_{sam}(\omega) = \frac{-2\ln\left[|H(\omega)|\right]^S + \alpha_{air}(\omega) L_{eff} - \alpha_{mat}(\omega)\left[L_{eff} - L_{sam}\right]}{L_{sam}}. \quad (7)$$

Above, $L_{eff}$ is the total propagation length of the THz beam. Albeit more complicated, we see that **Eqns. (6)** and **(7)** reduce to **Eqns. (2)** and **(3)** in the limit that $L_{eff} = L_{sam}$. The (S) superscript denotes operation on the sample. As done previously, the low power scan denotes the linear operation while the ensuing high-power scans elucidate on the nonlinear behavior. The full phase is represented as:

$$\Delta\phi(\omega) = \frac{\omega}{c}\left[\left(n_d^o(\omega) + n_{2,d}(\omega) I_{THz}\right) L_{eff} + \left(n_k^o(\omega) + n_{2,k}(\omega) I_{THz}\right) L_{sam}\right] \quad (8)$$

and the magnitude shift is represented as:

$$\Delta\ln\left[|H(\omega)|\right] = -\frac{1}{2}\left[\left(\alpha_d^o(\omega) + \alpha_{2,d}(\omega) I_{THz}\right) L_{eff} + \left(\alpha_k^o(\omega) + \alpha_{2,k}(\omega) I_{THz}\right) L_{sam}\right]. \quad (9)$$

A fit to **Eqns. (8)** and **(9)** solves for the value of $n_{2,k}(\omega) = n_{2,sam}(\omega) - n_{2,mat}(\omega)$. Finally, addition between $n_{2,d}$ and $n_{2,k}$ leads to $\Delta n_2 = n_{2,sam} - n_{2,air}$. Note that $n_k^o(\omega) = n_{sam}^o(\omega) - n_{mat}^o(\omega)$ and $\alpha$ follows respectively.

An analysis along the detectable bandwidth is made for liquid water, α-pinene, and Si. **Fig. 1(a)** plots the fit to the full phase in **Eqn. (8)** vs peak intensity for all three materials and the PLA material at 0.5 THz. In the presence of nonlinearity, the fit is a straight line with a slope proportional to the value of $\Delta n_2$. A non-zero slope indicates that there is nonlinear action present. **Fig. 1(b)** showcases the magnitude shift as a fit for **Eqn. (9)** vs peak intensity for all materials at 0.5 THz. The deviation of the liquid water magnitude in **Fig. 1(b)** is due to the strong absorption from the water sample while the high error on the first point of the liquid water phase fit in **Fig. 1(a)** is due to phase noise errors. The error bars indicate the phase-noise and fluctuation of the magnitude values. The phase noise is large because the liquid water attenuation is high, and the sample is relatively thin. Additionally, the collection optics after the sample position vignette some of the THz beam, further reducing the collected THz energy. Aside from these points, extracted phases and magnitudes agree very well with their fits and showcase that the values are nominally above the phase and magnitude noise. However, as can be seen, the fluctuations themselves do not follow a perfectly linear trend although they do collectively tend toward a straight line. The oscillation about the linear fit is due to stage error.

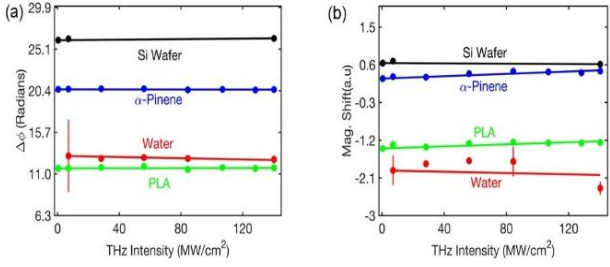

**Fig. 1.** Fitting lines performed in accordance to phase and magnitude equations for all tested materials at 0.5 THz. (a) Fit to **Eqn. (8)** vs peak intensity shows good agreement with data. Error bars indicate phase noise error. (b) Fitting lines to **Eqn. (9)** with respect to intensity. Error bars indicate standard deviations in the magnitude of the transfer function. The liquid water shows deviation in the magnitude due to high absorption.

The application of this analysis for every point in frequency along the detectable bandwidth (0.3-1.1 THz) for all materials yields the dispersion of $\Delta n_2$ and $\Delta \alpha_2$ in the samples. The dispersion of $\Delta n_2$ for the PLA material is shown in Fig. 2. The plot shows clear resonances corresponding to known spectral features for water vapor at 0.4, 0.56, 0.74, and 1.10 THz [14]. Because the orders of magnitude of the PLA $n_2$ and water vapor $n_2$ are expected to be the same, spectrally resolved Kerr coefficients are difficult to extract. Additionally, **Fig. 2** shows that as the nonlinear index changes its sign as the absorption reaches a peak value. This is consistent with the expected behavior from the Lorentz oscillator model [12]. Extension of the measurement window and a reduction of our time-constant will allow for a spectrally resolved demonstration of $n_2$ and $\alpha_2$ in the THz range.

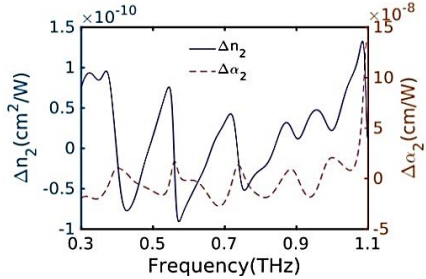

**Fig. 2.** Dispersion of the calculated nonlinear index difference as a result of applying **Eqns. (2)** and **(3)** for every point along the detectable frequency of PLA. The peaks coincide with known resonances for water vapor at 0.4, 0.56, 0.74, and 1.10 THz.

All samples are treated as non-centrosymmetric in our analysis based on their lack of molecular point symmetry. As such, the nonlinear absorption resonances can coincide with the linear resonances. According to the model presented in Ref. [1], while the main driving nonlinear process in the optical regime is governed by the electronic response of the third order susceptibility, it is theorized that the THz response is driven by molecular nonlinearities—that is to say, due to the rotational and vibrational modes of a molecule. Moreover, the presence of sharp absorption peaks for many materials in the THz regime allows for large values of $n_2$ given that the nonlinearity is expressed as a combination of off-resonant and resonant responses.

In the specific case of the PLA material, no sharp resonances were found in the linear regime evaluation, but the contribution of water vapor could not be decoupled from the contribution of the PLA. Fortunately, the differences between the PLA and water vapor $n_2$ as well as between sample and PLA $n_2$ are more important for determining the pure sample $n_2$. While the water vapor prevented a spectrally resolved measurement of $n_2$ and $\alpha_2$, spectral averaging upon the detectable bandwidth can be done and estimated values for the Kerr coefficients could be found. First, we acknowledge that Si is effectively a centrosymmetric molecule but may exhibit a non-centrosymmetric response due to its surface as is found in the optical regime. This effect would be very small and difficult to detect with our system. In our experiment, the peak field used to test Si is 230 kV/cm. The mean change to the index observed is $\Delta n = n_2 I = +0.008$. This sets a lower limit of detection for $n_2$ as $10^{-11}$ cm$^2$/W. Fortunately, the spectrally averaged value of Si has been evaluated before and found to be one order lower [15]. As such, the $\Delta n_2$ found in our experiment completely describes the spectrally averaged value of water vapor as $6\times10^{-11}$ cm$^2$/W. This value is in line with observations presented in Ref. [16].

Using this value as the mean reference and evaluating the mean nonlinear index differences, we can extract nonlinear index data for all other samples. The extracted values are listed in **Table 1**. The spectrally averaged values are shown for $n_2$ and $\alpha_2$ along 0.3-1.1 THz. The maximum THz peak field used is 230 kV/cm and the minimum peak field is 12.6 kV/cm. Linear refractive index values are shown for 0.9 THz and gathered at the lowest peak electric field. We anticipate that without eliminating the water vapor altogether, spectrally resolved Kerr coefficients cannot be easily extracted as known vapor resonances along 0.40, 0.56, 0.74, and 1.10 THz can cause considerable fluctuations [14]. These resonances also appear in the dispersion of $\Delta n_2$ as shown in **Fig. 2**.

**Table 1. Extracted Kerr Coefficients**

| Sample | $n$ | $\alpha$ (cm$^{-1}$) | $n_2$ (cm$^2$/W) | $\alpha_2$ (cm/W) |
|---|---|---|---|---|
| Si | 3.41 | 20 | -- | -- |
| H$_2$O Vapor | -- | -- | $0.6\times10^{-10}$ | $-2.0\times10^{-10}$ |
| Liquid H$_2$O | 2.33 | 230 | $7.8\pm3\times10^{-10}$ | $-2.0\times10^{-10}$ |
| α-pinene | 1.66 | 1.5 | $1.3\pm0.01\times10^{-10}$ | $-2.4\pm0.05\times10^{-10}$ |

Considering the model presented in [1], $n_2$ for liquid water can be as large as $7\times10^{-10}$ cm$^2$/W [4]. In our method, unlike the vapor phase, liquid water does not have sharp resonances within our detectable bandwidth. This means that the nonlinear coefficients are not expected to switch sign. With a peak field of 230 kV/cm and a mean phase-noise leading to $\Delta n_{noise} = +0.04$, liquid water has an approximate $n_2$ of $7.8\pm3\times10^{-10}$ cm$^2$/W. The high error is expected as the attenuation is strong in liquid water. Fortunately, the value is within the expected range. Lastly, α-pinene has very negligible absorption and dispersion along the linear regime and α-pinene can best be identified as having a C$_1$ point group molecular structure [17]. With our method, with a peak field of 230 kV/cm and a mean phase noise leading to $\Delta n_{noise} = +0.0002$, α-pinene has an $n_2$ of $1.30\pm0.01\times10^{-10}$ cm$^2$/W.

A flip in the sign for all samples was enacted via a phenomenological approach since physically, according to the model in Ref. [1], the mean Kerr coefficients for non-centrosymmetric samples are generally positive. A similar treatment for the nonlinear absorption reveals that $\alpha_2$ produces mean values of $-2.0\times10^{-10}$ cm/W for water vapor, $-4.4\pm0.2\times10^{-10}$ cm/W for liquid water, and $-2.40\pm0.05\times10^{-10}$ cm/W for α-pinene. These numbers indicate a mostly dominant saturable absorption process in all three samples. Specifically, for water, this observation was also seen in the Z-scan curves in Ref. [4]. The above result is interesting because it implies that a dynamic gain process may be possible for THz pulses inside the materials under strong peak fields (> 3 MV/cm). However, it is also possible that the contributions from the higher order odd nonlinearities may offset the reduction of the losses at high intensity. More detailed studies regarding spectrally resolved measurements will be required to confirm these claims. Moreover, our current results suggest that significant changes in the index of refraction are possible with viably reachable peak electric field strengths. For example, in the case of α-pinene, the linear index of refraction is 1.66 and a peak THz electric field of 2.2 MV/cm would induce a nonlinear index shift $\Delta n = 1.66$. At higher peak electric fields, the perturbative model for the nonlinear susceptibility is broken and a new formulation for the nonlinear index of refraction and nonlinear absorption must be considered. In such a regime, it is possible for the nonlinear index and nonlinear absorption to quickly saturate. This is important as it indicates that non-perturbative nonlinear optics can be studied in the THz regime without invoking relativistic laser intensities.

In closing, we experimentally demonstrate the spectrally averaged nonlinear Kerr coefficients for three samples (water, α-pinene, and Si) and water vapor along the spectral region between 0.3 to 1.1 THz. The work takes advantage of the fact that THz-TDS allows for a full-phase analysis, leading to direct methods of extracting nonlinear parameters from an experiment. The results show agreement with Z-scan techniques and are advantageous in that the beam mode need not be pre-specified. Furthermore, we plan to conduct these experiments in a humidity-controlled environment in order to fully and properly deduce the nonlinear coefficients in a spectrally resolved measurement. Future experiments will also focus on the study of the non-perturbative nonlinear optics in THz frequencies where relativistic laser intensities are not required.

**Funding.**

The research at University of Rochester is sponsored by the Army Research Office under Grant No. W911NF-17-1-0428, Air Force Office of Scientific Research under Grant No. FA9550-18-1-0357 and National Science Foundation under Grant No. ECCS-1916068. Kareem Garriga is also funded by the Gates Millennium Scholars Program.

**Acknowledgment.** The authors would like to recognize Prof. Jianming Dai, Prof. Govind Agrawal, and Greg Jenkins for fruitful discussions. The authors would also like to thank Dr. Peter Jepsen for his helpful suggestions, insight, and discussion.

**Disclosures**. The authors declare no conflicts of interest.